\documentclass[pre,twocolumn,showpacs,superscriptaddress,floatfix]{revtex4}
\usepackage{graphicx}

\newcommand{\be}{\begin{equation}}
\newcommand{\ee}{\end{equation}}
\newcommand{\bea}{\begin{eqnarray}}
\newcommand{\eea}{\end{eqnarray}}

\begin{document}

 \title{Aggregate formation in a system of coagulating and fragmenting
particles with mass-dependent diffusion rates}
 \author{R. Rajesh}
 \email{r.ravindran1@physics.ox.ac.uk}
 \affiliation{Department of Physics--Theoretical Physics, University of
Oxford, 1 Keble Road, Oxford OX1 3NP, UK}
 \author{Dibyendu Das}
 \affiliation{Martin Fisher School of Physics, Brandeis University, 415
South Street, Waltham, MA 02454, USA}
 \author{Bulbul Chakraborty}
 \affiliation{Martin Fisher School of Physics, Brandeis University, 415
South Street, Waltham, MA 02454, USA}
 \author{Mustansir Barma}
 \affiliation{Department of Theoretical Physics, Tata Institute of
Fundamental Research, Homi Bhabha Road, Mumbai-400005, India}

\date{\today}

\begin{abstract}
 The effect of introducing a mass dependent diffusion rate $\sim
m^{-\alpha}$ in a model of coagulation with single-particle break up is
studied both analytically and numerically.  The model with $\alpha=0$ is
known to undergo a nonequilibrium phase transition as the mass density in
the system is varied from a phase with an exponential distribution of mass
to a phase with a power-law distribution of masses in addition to a single
infinite aggregate. This transition is shown to be curbed, at finite
densities, for all $\alpha > 0$ in any dimension. However, a signature of
this transition is seen in finite systems in the form of a large aggregate
and the finite size scaling implications of this are characterized. The
exponents characterizing the steady state probability that a randomly
chosen site has mass $m$ are calculated using scaling arguments. The full
probability distribution is obtained within a mean field approximation and
found to compare well with the results from numerical simulations in one
dimension. 
 \end{abstract}


\maketitle

\section{Introduction}

Systems far from equilibrium can undergo phase transitions between two
types of steady states when the parameters of the system are varied. It is
important to ask about the sensitivity of such nonequilibrium phase
transitions to changes in the governing dynamics. If the transition
survives, is the universality class affected? If the transition is lost,
does a signature of the lost phase remain in any form? 

In this paper, we investigate these questions within a lattice model of
coagulation and fragmentation in which the diffusion constant for a mass
$m$ varies as $m^{-\alpha}$ with $\alpha > 0$. For the case in which
diffusion is independent of the mass $(\alpha = 0)$ and fragmentation
involves only chipping off of unit masses, it is known that there is a
phase transition from a low-density phase with an exponential distribution
of masses to a high-density phase with a power-law distribution of masses
in addition to an infinite aggregate with a mass proportional to the
volume $V$ \cite{MKB}. This transition is characterized by a new
universality class, different from familiar classes such as directed
percolation or the parity-conserving class \cite{hinrichsen}, wetting
transitions, roughening transitions or boundary-driven transitions
\cite{mukamel}. We will show below that this high-density phase is lost as
soon as $\alpha$ is nonzero. Remarkably, though, an imprint of the
infinite aggregate remains in the form of a large aggregate that strongly
modifies the finite-size behavior of the system, and we characterize the
scaling implications of this. 

Let us summarize the results of earlier related work. Enhancement of
aggregation moves with increasing mass, corresponding to negative values
of $\alpha$, have been investigated earlier in the context of coalescing
branched polymers. For $\alpha = -1$, using a Smoluchowski approach it was
shown that the system undergoes a gelation transition, $i.e.$, an
aggregate that subsumes a finite fraction of the total mass forms at
finite time.  The fragmentation move was shown to modify the mass
distribution power law exponent at the gelation transition \cite{VZL}. An
off-lattice version of the $\alpha=0$ case was studied \cite{KR} using
Smoluchowski rate equations in the context of aggregation in dry
environments.  In these studies \cite{MKB, VZL, KR}, the coagulating and
fragmenting masses represented polymers in a solution, undergoing
polymerization and depolymerization. In a realistic situation, it may be
expected that the diffusion of the polymers would depend on their masses.
For example, in the well known models of polymer motion such as the Rouse
model or the Zimm model \cite{edwards}, the polymer diffusion constant
$D(m) \sim m^{-1}$ and $m^{-1/2}$ respectively. This would correspond to
$\alpha = 1$ and $1/2$ in our model. This provides a further motivation
for studying the model with a mass-dependent diffusion rate. 

Other modifications of the dynamics of the $\alpha=0$ model that have been
studied include changes of the fragmentation rule, the introduction of a
spatial bias in the dynamics, and the effects of quenched disorder. 
Introduction of a mass-dependent fragmentation by allowing fractions of
masses to break off (as opposed to single-particle break up) was studied
in \cite{RM2,KG}. In this case, it could be inferred that the phase
transition is curbed in all dimensions.  Spatial bias was introduced by
choosing rates such that masses have a preferred direction of motion, but
with mass-independent hopping rates. In this case, it was shown that the
phase transition is curbed in one dimension \cite{RK}. In two and higher
dimensions, it was shown that bias is irrelevant at least as far as the
existence of a phase transition was concerned.  Finally, in a disordered
model where fragmentation of masses could occur only at fixed sites, it
was shown that even in the limit of very low disorder, a new mechanism for
the formation of localized infinite aggregates sets in \cite{KJ}. 

The remainder of the paper is organized as follows. Section II contains
the definition of the model, a brief review of earlier results and a
summary of results obtained in this paper. Section III contains the
analytical proof for the nonexistence of a phase with an infinite
aggregate at large densities for any non-zero value of $\alpha$. In Sec.
IV, the exponents associated with the probability distribution $P(m)$ are
determined using scaling arguments.  Results of Monte Carlo simulations in
one dimension are also presented. In Sec. V, the full distribution is
obtained from a mean field approximation and compared with the $P(m)$
obtained from numerical simulations. The appendix discusses different
limiting cases of the problem that are solvable exactly. 

\section{\label{sec2}Model and Results}

\subsection{\label{sec2a} The model}

The model is defined on a $d$-dimensional hyper-cubic lattice with
periodic boundary conditions.  Starting from a random distribution of
non-negative integer masses at each site, the system evolves in time via
the following microscopic moves : (1) each mass $m$ hops with rate $D(m)=
m^{-\alpha}$ to one of its nearest neighbor sites chosen randomly (2) with
rate $w$, unit mass breaks off from an already existing mass and is
transferred to a randomly chosen neighboring site and (3) following moves
(1) and (2), the mass at each site adds up. The mass density $\rho$ is a
conserved quantity in the model.

In one dimension, this model can be mapped \cite{MKB} onto other well
studied models of nonequilibrium statistical mechanics. By interpreting
the masses as interparticle spacings, the model is mapped onto a
one-dimensional hard core lattice gas model with competing short and long
range hops.  Correspondingly, the problem may be mapped onto a fluctuating
interface with competing short and long range moves.  The limiting case
$w=\infty$ reduces to the well studied simple exclusion process
\cite{evans} or equivalently to a fluctuating interface governed by the
Edwards-Wilkinson equation \cite{EW}.

\subsection{\label{sec2b}Previous results for $\alpha=0$}

The case $\alpha=0$ was studied by means of a mean field approximation
\cite{MKB}, analytical calculations \cite{RM1} and numerical simulations
in \cite{MKB,RM1}. The results are summarized below. The steady state
single-site mass distribution $P(m)$ was shown to undergo a phase
transition in all dimensions. In the $\rho - w$ plane, there is a critical
line $\rho_c(w) = \sqrt{1+w}-1$ that separates two types of asymptotic
behavior of $P(m)$.  For fixed $w$, as $\rho$ is varied across the
critical line $\rho_c(w)$, the large $m$ behavior of $P(m)$ was shown to
be
 \be
 P(m) \sim \cases{
 e^{-m/m^*}, & $\rho< \rho_c(w)$,\cr
 m^{-\tau}, &$\rho=\rho_c$,\cr
 m^{-\tau} +\mbox{infinite aggregate,}&$\rho>\rho_c(w)$,\cr}
 \label{eq:1}
 \ee
 where by ``infinite aggregate'', we mean a cluster that contains a finite
fraction of the total mass in the system.  That is, the tail of the mass
distribution changes from an exponential decay to an algebraic one as
$\rho$ approaches $\rho_c$ from below. As one increases $\rho$ beyond
$\rho_c$, the asymptotic algebraic part of the critical distribution
remains unchanged but in addition an infinite aggregate forms. All the
additional mass in excess of the critical mass condenses into this single
cluster and does not disturb the background critical distribution. The
mathematical mechanism giving rise to the formation of the infinite
aggregate at the onset of the phase transition was found to be very
similar to that of the equilibrium Bose-Einstein condensation in an ideal
Bose gas. 

Finite size effects in the aggregate phase were studied in \cite{RM1}. For
a system of size $V$, the probability distribution $P(m, V)$ for $\rho
\geq \rho_c$ was assumed to have the scaling form
 \be
 P(m,V)\approx \frac{1}{m^\tau} f\left(\frac{m}{V^\chi}\right)+\frac{1}{V}
\delta \left[ m- (\rho-\rho_c)V \right],
 \label{eq:2}
 \ee
 where the exponent $\chi$ is a crossover exponent, and the $\delta$
function indicates the aggregate part. The exponents $\chi$ and $\tau$
were shown to be related by the scaling relation $\chi (\tau - 1)=1$.  The
exponent $\tau$ was shown to be $5/2$ in the mean field approximation
\cite{MKB}; further, numerical evidence was presented \cite{RM1} for the
exponent being the same in all dimensions. 

\subsection{\label{sec2c}Summary of new results in this paper}

The principal results obtained in this paper are summarized below. 

\noindent (i) It is shown analytically that there is no phase transition
at finite density for any $\alpha>0$ in any dimension.

\noindent (ii) On an infinite lattice with fixed density $\rho$, on
assuming a scaling form
 \be
 P(m,\rho) = \frac{1}{m^{\tau'}} f\left(\frac{m}{\rho^\phi}\right),
 \label{eq:3}
 \ee 
 where $f(y)$ falls exponentially as $y \to \infty$, it is shown that the
two exponents are related to each other by the scaling relation
 \be
 \phi(2-\tau')=1. 
 \label{eq:4}
 \ee 
 The power law exponent $\tau'$ is shown to be equal to
 \be
 \tau' = \cases{
 2-\frac{\alpha}{2} &for $0<\alpha \leq 2$,\cr
 1 &for $\alpha>2$.\cr}
 \label{eq:5}
 \ee 
 Equivalently,
 \be
 \phi = \cases{\frac{2}{\alpha} &for $0<\alpha \leq 2$,\cr
 1 &for $\alpha > 2$. \cr} \label{eq:6}
 \ee

\noindent (iii) In numerical simulations on a finite one dimensional
lattice, it is seen that an aggregate forms when the total mass in the
system is increased beyond a certain critical value. By analogy with the
$\alpha =0$ case, we make the assumption that $P(m)$ has the scaling form,
 \be
 P(m,V)\approx \frac{1}{m^{\tau}}
 g\left(\frac{m}{V^\chi}\right)+\frac{1}{V} \delta(m- M-M_c),
 \label{eq:7}
 \ee 
 where $M_c$ is a $V$ dependent critical mass. It is argued that
$\tau'=\tau$ with $\chi$ being related to $\tau$ through $\chi (\tau - 1)
= 1$ as in the $\alpha=0$ case. The critical mass is shown to scale with
system size as
 \be
 M_c\sim V^{2/(2-\alpha)}, \quad \mbox{for}~\alpha<2, 
 \label{eq:8}
 \ee
 implying that the critical density $\rho_c = M_c / V$ diverges with the
system size.

\noindent (iv) By means of a mean field approximation, we obtain the full
probability distribution $P(m)$. The scaling form Eq.~(\ref{eq:3}) is seen
to hold with the exponents as given in Eqs.~(\ref{eq:5}) and (\ref{eq:6}).

\section{\label{sec3}Arguments for no phase transition at finite density
for $\alpha>0$}

\begin{figure}
 \includegraphics[width=8.0cm]{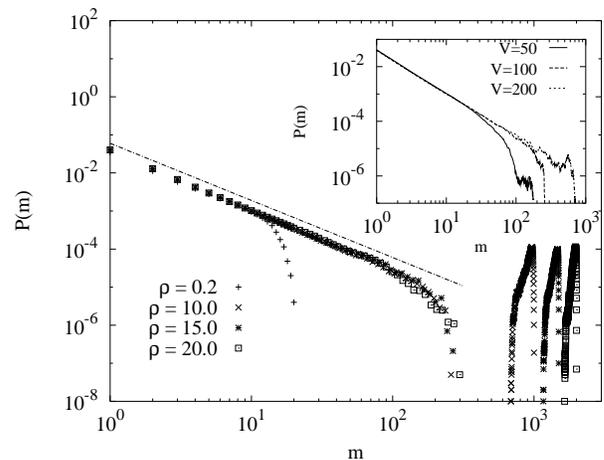}
 \caption{\label{fig1}The variation of $P(m)$ with $m$ for $\alpha = 1.0$
is shown for four different values of density $\rho$ at fixed lattice size
$V$. As density is increased, $P(m)$ changes from an exponential
distribution to a power law distribution. On increasing $\rho$ further,
the power law part remains unchanged while the mass in excess of a
critical density $\rho_c$ condenses into an aggregate. The straight line
has a power $-1.5$. The simulation results are for a one-dimensional
lattice of size $100$ and $w=0.1$. In the inset, the variation of the
power law cutoff with system size is shown. The simulation results are
for a one-dimensional lattice with $\rho=10.0$ and $w=0.1$.}
 \end{figure}

On a finite lattice, on increasing the total mass $M$ from zero to large
values, the following behavior is observed in numerical simulations. For
small values of $M$, $P(m)$ is seen to have an exponential tail for large
mass (see Fig.~\ref{fig1}). As $M$ is increased to a critical value $M_c$,
$P(m)$ changes to a power law with a cut off at large $m$. As $M$ is
increased beyond $M_c$, an aggregate forms that contains all the mass in
excess of $M_c$. The rest of the distribution remains identical to the one
at $M_c$. The power law part has a lattice size dependent cutoff (see
inset of Fig.~\ref{fig1}). All these observations are qualitatively
similar to the $\alpha =0$ case. A crucial difference is the fact that the
power law exponent is seen to be less than $2.0$ for $\alpha>0$. This is a
puzzle since finite density would imply that $\tau>2$. In this section, we
prove that there is in fact no transition at finite densities in the
thermodynamic limit. The transition seen in finite size simulations is
explained by the fact that $M_c$ no longer scales as $V$ (as in the
$\alpha=0$ case), but with a power of $V$ greater than unity.

We show that an aggregate with $M_c \propto V$ cannot be stable at finite
densities by assuming the presence of such an aggregate and showing that
this leads to a contradiction.  In Sec.~\ref{sec3a}, we study the mass
profile as a function of distance from the aggregate. Based on our
observation that at distances far from the aggregate the state of the
system resembles that at the transition point, we obtain exact relations
that the critical point should satisfy.  In Sec.~\ref{sec3b}, we derive
further exact relations by examining the two point correlations. In
Sec.~\ref{sec3c}, we show that the relations obtained from
Secs.~\ref{sec3a} and \ref{sec3b}, when put together, imply that there can
be no phase transition at finite densities.

\subsection{\label{sec3a}Reference frame fixed to the aggregate}

In the aggregate phase of the $\alpha = 0$ model, it is known that there
exists only one large aggregate \cite{RM1} in steady state; if there were
more than one, they would collide and coalesce into one.  This scenario is
verified as well in numerical simulations for arbitrary $\alpha$ (the area
under the aggregate part in the mass distribution being equal to $1/V$).
Further, in the limit $V\rightarrow \infty$, the aggregate becomes
immobile for $\alpha>0$ because its mass diverges with system size. 

Consider a frame of reference that is attached to this aggregate.  Let
$m_{\bf x}$ and $s_{\bf x}$ denote the mass and occupation probability at
a site ${\bf x}$ with respect to the aggregate. Then, by examining the
inflow and outflow of mass at each site, we obtain
 \bea
 \frac {d \langle m_{\bf x} \rangle}{dt} &=& - \left[ w s_{{\bf x}} +
\langle m_{{\bf x}}^{1-\alpha}\rangle (1-\delta_{{\bf x},{\bf 0}}) \right]
\nonumber \\
 && \mbox{} + \frac{1}{2 d} \sum_{{\bf x}^{\prime}} \left( w s_{{\bf
x}^{\prime}}+ \langle m_{{\bf x}^{\prime}}^{1-\alpha}\rangle
 \right),
 \label{eq:9}
 \eea
 with $s_{{\bf 0}}=1$ and $\langle m^y \rangle = \sum_{m=1} P(m) m^y$.  In
the steady state, the time derivative is set to zero. Then, the solution
of Eq.~(\ref{eq:9}) is
 \be
 \langle m_{{\bf x}}^{1-\alpha} \rangle + w s_{{\bf x}} = w \quad
\mbox{for}~{\bf x} \neq 0. 
 \label{eq:10}
 \ee
 At distances far away from the aggregate, the state of the system
resembles that at criticality. Taking the limit $|{\bf x}| \rightarrow
\infty$ in Eq.~(\ref{eq:10}), we obtain
 \be
 {\langle m^{1-\alpha} \rangle}_c = w (1-s_c) \quad \mbox{for}~ \alpha >0. 
 \label{eq:11}
 \ee
 This is a relation that the system should satisfy at the critical point. 
 
In the case $\alpha=0$, the aggregate is mobile. When the aggregate hops,
this corresponds to all the other particles simultaneously making a hop
with respect to the aggregate. An analysis, similar to the one carried out
for $\alpha >0$, yields
 \be
 2 \rho_c = w (1- s_c) \quad \mbox{for}~\alpha=0. 
 \label{eq:12}
 \ee
 The origin of the factor $2$ may be traced to the fact that the aggregate
is mobile.

\subsection{\label{sec3b}Two point correlations}

In this section, we derive further exact relations that the system
satisfies at the critical point by studying the two-point correlations.
The analysis is similar to the analysis done for the $\alpha=0$ case
\cite{RM1}. In the rest of the paper, we will work in a coordinate system
fixed to an arbitrary fixed site. To fix notation, ${\bf x}'$ will always
denote one of the $2d$ nearest neighbors of ${\bf x}$, while ${\bf x}_o$
will denote a neighbor of the origin ${\bf 0}$. Let $\eta({\bf x},{\bf
x}',t)$ be the mass transferred from site ${\bf x}$ to ${\bf x}'$ at time
$t$ in a time interval $\Delta t$. From the definition of the model, it
follows that
 \be
 \eta({\bf x},{\bf x}',t) = \cases{
 m_{{\bf x}} & with prob. $\frac{1}{2 d}\frac{\Delta t}{m_{{\bf
x}}^{\alpha}}$ \cr
 1-\delta_{m_{{\bf x}},0} &with prob. $\frac{1}{2 d} w {\Delta t}$ \cr
 0 & otherwise. \cr}
 \label{eq:13}
 \ee
 To order $\Delta t$, the only nonzero two point correlation in the noise
is
 \be
 \langle \eta({\bf x}_1,{\bf x}^\prime_{1})^2\rangle = \frac{\Delta t}{2
d} \left[m_{{\bf x}_1}^{2-\alpha}+ w (1-\delta_{m_{{\bf x}_1},0}) \right]. 
 \label{eq:14}
 \ee

The mass $m_{{\bf x}}(t)$ at lattice site ${\bf x}$ at time $t$ evolves as
 \be
 m_{{\bf x}}(t+{\Delta t}) = m_{{\bf x}}(t) - \sum_{{\bf x}'} \eta({\bf
x},{\bf x}',t) + \sum_{{\bf x}'}\eta({\bf x}',{\bf x},t). 
 \label{eq:15}
 \ee
 To obtain the two point correlations, we multiply $m_{{\bf x}}(t+{\Delta
t})$ by $m_{{\bf 0}}(t+{\Delta t})$ and take averages over the possible
stochastic moves and then over the steady state ensemble of states.
Dropping all time derivatives and using Eqs.~(\ref{eq:13}), (\ref{eq:14}) 
and (\ref{eq:15}), we obtain
 \bea
 \lefteqn{C_{\alpha}({\bf x}) - w D({\bf x})-\frac{1}{2 d} \sum_{{\bf x}'}
\left[C_{\alpha}({\bf x}') -w D({\bf x}') \right]} \nonumber\\
 && \mbox{} = \left[C_{\alpha}({\bf 0}) + w s \right] \left(\delta_{{\bf
x},{\bf 0}} - \frac{1}{2 d}\sum_{{\bf x}_o} \delta_{{\bf x},{\bf x}_o}
\right),
 \label{eq:16}
 \eea
 where $C_{\alpha}({\bf x})=\langle m_{{\bf x}}m_{{\bf 0}}^{1-\alpha}
\rangle$ and $D({\bf x}) = \langle m_{{\bf x}} \delta_{m_{{\bf 0}},0}
\rangle$. The homogeneous part of Eq.~(\ref{eq:16}) is the Laplace
equation ${\nabla^{2}} \left[C_{\alpha}({\bf x}) - wD({\bf x})\right] =
0$. With the inhomogeneous part, the unique solution is
 \be
 C_{\alpha}({\bf x}) = w[D({\bf x}) - s] \quad \mbox{for} ~{\bf x} \neq
{\bf 0}. 
 \label{eq:17}
 \ee
 Equation~(\ref{eq:17}) is a relation between two point correlations. A
relation between one point functions is obtained by summing over all ${\bf
x}$, the simplifying factor being that total mass is conserved. Thus,
 \be
 M \langle m^{1-\alpha} \rangle - \langle m^{2-\alpha} \rangle = w M (1-s) 
- w s V + w s.
 \label{eq:18}
 \ee
 This is an exact relation in all dimensions. 

We are interested in the limit when $M,V \rightarrow \infty$ keeping the
density $\rho$ fixed. Taking this limit in Eq.~(\ref{eq:18}), we obtain
 \be
 \rho {\langle m^{1-\alpha} \rangle} - \frac{\langle
m^{2-\alpha}\rangle}{V} = w \rho (1-s) -w s, \quad V \gg 1.
 \label{eq:19}
 \ee
 In the exponential phase, $\langle m^{2-\alpha} \rangle$ is finite and
hence $\langle m^{2-\alpha} \rangle/V \rightarrow 0$ as $V\rightarrow
\infty$. At the transition point and in the aggregate phase $\langle
m^{2-\alpha} \rangle$ can at most diverge as $V^{1-\alpha}$ (cf. 
discussions in the later sections of this paper). This implies that
$\langle m^{2-\alpha} \rangle/V \rightarrow 0$ as $V\rightarrow \infty$
for all finite densities and any $\alpha > 0$.  Thus, another exact
relation at the critical point is obtained:
 \be
 \rho_c {\langle m^{1-\alpha} \rangle}_c = w \rho_c (1-s_c) -w s_c. 
 \label{eq:20}
 \ee

\subsection{\label{sec3c}Proof of no transition}

We combine the results of Secs.~\ref{sec3a} and \ref{sec3b} to show that
there is no transition. The three quantities $\langle m^{1-\alpha}
\rangle_c$, $\rho_c$ and $s_c$ have to simultaneously satisfy two
relations, namely Eqs.~(\ref{eq:11}) and (\ref{eq:20}). For non-zero
values of $\alpha$, this is possible only when either $w=0$ and $\rho_c=0$
or $\rho_c=\infty$. Equations~(\ref{eq:11}) and (\ref{eq:20}) cannot be
satisfied at finite nonzero values of $\rho_c$.  This completes the proof
that there is no transition for $\alpha>0$ at finite critical density
$\rho_c$. 

As a check of correctness, $\rho_c$ and $s_c$ can be calculated for the
$\alpha=0$ case from Eqs.~(\ref{eq:12}) and (\ref{eq:20}). We obtain
$\rho_c(w) =\sqrt{1+w}-1$ and $s_c= (w+2-2 \sqrt{1+w})/w$. Not
surprisingly, this is the result that had been obtained in \cite{RM1} for
the $\alpha=0$ case.

The fact remains that a single large aggregate is seen in simulations on a
finite lattice when the mass is large enough (see Fig.~\ref{fig1}).  This
observation would be consistent with the above result that there is no
transition, provided the critical density $\rho_c$ seen in simulations 
diverges with $V$ as $V^\beta$ with $\beta>0$. We address this in
the next section. 

\section{\label{sec4}Scaling forms for probability distribution
$P(m,\rho,V)$}

\subsection{\label{sec4a} Large finite densities}

In this subsection, we consider the case when the total mass $M$ and the
system size $V$ are increased to infinity keeping the density $\rho=M/V$
fixed. In this case, the system is always in the exponential phase.  We
assume the following scaling form for the probability distribution: 
 \be
 \lim_{V\rightarrow \infty} P(m,V,\rho) \sim \frac{1}{m^{\tau'}}
f_{\alpha} \left( \frac{m}{\rho^\phi} \right),
 \label{eq:21}
 \ee
 where $\tau'$ and $\phi$ are two unknown exponents.  Rigorous upper and
lower bounds can be placed on $\tau'$. Clearly, $\langle m \rangle = \int
d m ~m P(m,V,\rho)$ should diverge as $\rho$ when $\rho \rightarrow
\infty$. But, the different moments of $m$ vary with $\rho$ as
 \be \int\! d m ~m^y P(m, \rho) = \int \! d m ~m^{y-\tau'} f_{\alpha}
\left( \frac{m}{\rho^\phi} \right) \sim \rho^{\phi (1+y-\tau')}.
 \label{eq:22}
 \ee
 This implies that $\tau' \leq 2$.  Also, from Eq.~(\ref{eq:19}), $\langle
m^{1-\alpha} \rangle$ is seen to be finite for all $\rho$, in particular
for $\rho \rightarrow \infty$. This implies that $\tau' > 2-\alpha$. Also,
from the requirement that probability distribution sums up to $1$, $\tau'$
necessarily has to be greater than $1$. These bounds can be summarized as
 \be
 \max (2-\alpha,1) < \tau' \leq 2. 
 \label{eq:23}
 \ee

The two exponents $\tau'$ and $\phi$ can be expressed in terms of one
another by an exponent equality. The average mass $\langle m \rangle =
\rho$. This implies that
 \be
 \phi (2-\tau')=1. 
 \label{eq:24}
 \ee
 Thus, there is only one independent exponent. $\tau'$ is determined in
Sec.~\ref{sec4b} by studying the finite size corrections to the
probability distribution.
 
\subsection{\label{sec4b} Aggregate formation on large finite lattices}

For a system on a finite lattice, we see (Fig.~\ref{fig1}) that when the
total mass is increased beyond a critical mass $M_c(V)$, the probability
distribution has a $V$-dependent cutoff. Any additional mass that is added
aggregates together to form one massive aggregate. Using this information,
we assume the following form for the probability distribution: 
 \be
 P(m,V) = \frac{1}{m^{\tau}} g_{\alpha}(\frac{m}{V^{\chi}}) + \frac{1}{V}
\delta (m- (M-M_c)), \label{eq:25}
 \ee
 where $M$ is the total mass in the system. The two exponents $\tau$ and
$\chi$ can be expressed in terms of the two other exponents $\tau'$ and
$\phi$. We then determine $\tau$ by scaling arguments, thus solving for
all the exponents.

In \cite{RM1}, it was shown that $\tau$ and $\chi$ are related by the
scaling relation
 \be
 \chi (\tau-1) =1.  \label{eq:26}
 \ee
 The derivation of this result was based on the fact that the number of
aggregates is of order unity.  The arguments carry forward to the general
$\alpha$ case without any modification. We now argue that $\tau'=\tau$
from Eqs.~(\ref{eq:21}) and (\ref{eq:25}). The system feels the presence
of the finite size when the density dependent cutoff in Eq.~(\ref{eq:21}) 
becomes of the same order as the lattice size dependent cutoff in
Eq.~(\ref{eq:25}). That is, when $\rho_c^\phi \sim V^\chi$, or $\rho_c
\sim V^{\chi/\phi}$. But, $\rho_c$ is the mean value of the mass in the
power law part and from Eq.~(\ref{eq:25}), $\rho_c \sim V^{\chi(2-\tau)}$.
Thus,
 \be
 \chi (2-\tau) = \frac{\chi}{\phi}. 
 \label{eq:27}
 \ee
 Substituting for $\phi$ in terms of $\tau'$ (see Eq.~(\ref{eq:24})), we
obtain
 \be
 \tau'= \tau.  \label{eq:28}
 \ee
 That leaves only one undetermined exponent in terms of which all the
other exponents can be expressed. 

To determine this exponent, we start with Eq.~(\ref{eq:18}) at the
transition point, namely,
 \be
 M_c \langle m^{1-\alpha} \rangle_c - \langle m^{2-\alpha} \rangle_c = w
M_c (1-s_c) - w s_c V + w s_c.
 \label{eq:29}
 \ee
 Unlike the scaling $M/V=\rho$ that we used in deriving Eq.~(\ref{eq:19}) 
from Eq.~(\ref{eq:18}), we now assume that $M_c$ scales as some power of
$V$, namely $M_c \sim V^{\beta+1}$, with $\beta>0$. From
Eq.~(\ref{eq:25}), we obtain
 \be
 \beta= \chi (2-\tau). 
 \label{eq:30}
 \ee
 Firstly, by substituting Eq.~(\ref{eq:25}) in Eq.~(\ref{eq:18}), it is
easy to derive that, to leading order in $V$, $\langle m^{1-\alpha}
\rangle_c = w (1-s_c)$. Now, to satisfy Eq.~(\ref{eq:29}), there are two
cases we have to consider : (A)  $\langle m^{2-\alpha} \rangle_c \sim V$,
or (B) $\langle m^{2-\alpha} \rangle_c \sim \rm{const}$ \textit{and}
$\langle m^{1-\alpha} \rangle_c = w (1-s_c) - w s_c V^{-\beta} + \ldots$.
Case~(A) requires that $\chi(3-\alpha-\tau)=1$, which when simplified
implies that $\tau=2-\alpha/2$. Case~(B) requires that
$\chi(3-\alpha-\tau)<0$ and $\chi(2-\alpha-\tau) \leq -\beta$ which
implies that $\tau> 3-\alpha$ and $\tau\geq 2-\alpha/2$. For $\alpha\leq
1$, these bounds are in contradiction with the rigorous bounds
Eq.~(\ref{eq:23}). Thus for $0<\alpha<1$, only case~(A) is viable and
hence $\tau = 2-\alpha/2$. For $1<\alpha \leq 2$, we have to consider
case~(B) also. However, any solution that arises from choosing case~(B)
would imply a non monotonic dependence of $\tau$ on $\alpha$. However, we
expect that $\tau$ is a monotonic function of $\alpha$, and hence we
discard the solutions arising from case~(B). Thus, $\tau=2-\alpha/2$.  For
$\alpha>2$, this solution is in contradiction with the rigorous lower
bound Eq.~(\ref{eq:23}). Therefore, we assume that the exponent value is
stuck at $1$ for all $\alpha>2$ (There is no contradiction with the above
derivation since if the distribution were indeed a power law, then the
integrals would now diverge at the lower cutoff too). This agrees with the
exact solution of the $\alpha=\infty$ case (see Appendix) in which case
$\tau=1$. Thus,
 \be
 \tau= \cases{
 \frac{5}{2} &\mbox{for} $\alpha=0$,\cr
 2 - \frac{\alpha}{2} &\mbox{for} $0 <\alpha \leq 2$,\cr
 1 &\mbox{for} $\alpha >2$,\cr} \label{eq:31}
 \ee
 where the value for $\alpha = 0$ is from \cite{MKB,RM1}. Solving for the
other exponents, we obtain, for $0<\alpha<2$,
 \bea
 \chi &=& \frac{2}{2-\alpha}, \label{eq:32}\\
 \beta &=& \frac{\alpha}{2-\alpha}, \label{eq:33}\\
 \phi &=& \frac{2}{\alpha}. \label{eq:34}
 \eea

Now that all the exponents are known, we return to the behavior of the
scaling function associated with $P(m)$ at large finite densities. 
Numerically, we observe that the scaling function $f_{\alpha}(x) \sim
\mbox{const}$ as $x\rightarrow 0$ for $0<\alpha<2$. For $\alpha>2$, we
expect $f_{\alpha}(x)$ to go to zero as some power of $x$ as $x\rightarrow
0$ (see Sec.~\ref{sec5} for numerics). This means that, for $\alpha<2$, in
the limit $\rho\rightarrow \infty$, the probability distribution is a
power law despite the mean mass diverging. These observations are
consistent with the exact solution of the $\alpha=\infty$ case (see
Appendix).  The formation of a power law in the limit of $\rho\rightarrow
\infty$ is similar to observations in models of aggregation in the
presence of a constant influx of particles from outside
\cite{takayasu,KMR}.  In these models, despite the mean mass diverging
with time, $P(m)$ develops into a power law distribution.

An implication of the exponent $\tau$ being less than two is that the
average time scale in the system may become very large. The average time
scale goes as the average of the inverse of the diffusion constant, i.e.
$\langle 1/D(m) \rangle = \langle m^{\alpha} \rangle \sim m_{*}^{3\alpha/2
- 1}$ where $m_{*}$ is the mass cutoff $\sim \rho^{\phi}$. Thus for
$\alpha > 2/3$, it would diverge with $m_*$.  On the other hand, the
inverse of the average diffusion constant $1/{\langle D(m) \rangle}$
remain finite, since ${\langle D(m) \rangle} = \langle m^{-\alpha}
\rangle$ is always finite.  Thus our model produces a broad distribution
of time scales with dissonance of average of its inverse, and inverse of
its average. Such a scenario is reminiscent of diffusion in heterogeneous
environment which arises in super-cooled liquids \cite{tarjus}. In the
latter system, translational diffusion constant averaged over several
heterogeneous regions falls out of proportionality with inverse of the
average time scale. However, the connection of our model to the
super-cooled liquids should not be taken too seriously since while the
latter is in equilibrium, our model exhibits a nonequilibrium steady
state.

\subsection{\label{sec4c}Numerical checks}

\begin{figure}
 \includegraphics[width=8.0cm]{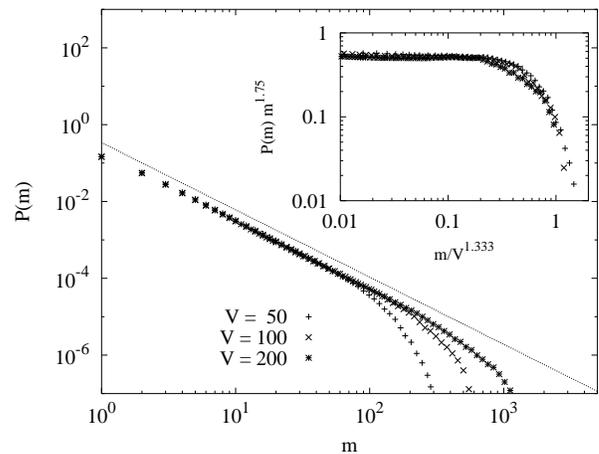}
 \caption{\label{fig2}The power law part of $P(m)$ obtained from Monte
Carlo simulations is shown for three different values of $V$. The
simulations are on a one dimensional lattice with $w=1.0$, $\rho=15.0$ and
$\alpha=0.5$. The straight line has an exponent $-1.75$ (see
Eq.~(\ref{eq:31})).  In the inset the scaling plot of these curves are
shown when scaled as in Eq.~(\ref{eq:25})}
 \end{figure}

In this subsection we provide numerical support for the assertions in
Sec.~\ref{sec4b}, from Monte Carlo simulations in one dimension. Due to
finite size effects, it is difficult to make an accurate direct
measurement of the exponent $\tau$ from Monte Carlo simulations for
$P(m)$. However, we show that the analytic results for the power law
exponents are consistent with the numerically obtained $P(m)$. In
Fig.~\ref{fig2}, the results from simulations are compared with the
analytic results for $\alpha=0.5$. In the inset, when the plots for
different $V$ are scaled as in Eq.~(\ref{eq:25}), the curves lie on top of
each other. For $\alpha=1.0$, the predicted exponent $1.5$ also matches
very well with simulations (see Fig.~\ref{fig1}). 
 \begin{figure}
 \includegraphics[width=8.0cm]{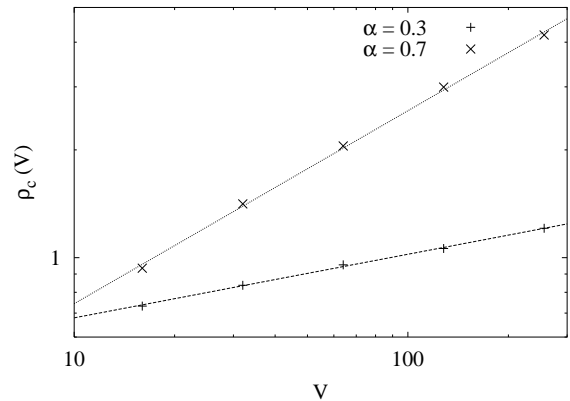}
 \caption{\label{fig3}The variation of $\rho_c(V)$ with $V$ is shown for
$\alpha=0.3$ and $\alpha=0.7$, where the upper curve has been shifted
downwards for clarity. The straight lines are best fit with power law
exponents equal to $0.178\pm 0.004$ for $\alpha=0.3$ and $0.539\pm 0.011$
for $\alpha = 0.7$. These values should be compared with the analytic
results $0.176\ldots$ for $\alpha=0.3$ and $0.538\ldots$ for $\alpha =0.7$
(Eq.~(\ref{eq:33})). The simulation was done on one-dimensional lattices
for $w=1.0$.}
 \end{figure}

As a second check, we measured $\rho_c(V)$ as a function of $V$ for
$\alpha=0.3$ and $\alpha=0.7$. We adopted the following procedure for
measuring $\rho_c(V)$. We start the system with total mass much greater
than the critical mass $\rho_c(V) V$. The system is allowed to reach the
steady state. The cluster with the largest mass is identified as the
infinite cluster. $\rho_c(V)$ is obtained by measuring the average mass in
the rest of the system (excluding the infinite aggregate).  In
Fig.~\ref{fig3}, we obtain the exponent $\beta$ from the slope of a
log-log plot of $\rho_c(V)$ versus $V$. There is excellent agreement with
the analytically predicted values. 

\section{\label{sec5}Mean field approximation}

In Sec.~\ref{sec4}, the exponents characterizing the probability
distribution $P(m)$ were calculated. These exponents were independent of
dimension and hence should match with the mean field exponents. Also, it
was observed \cite{RM1} in the $\alpha=0$ case that the mean field $P(m)$
matched very well with the numerically obtained $P(m)$ for all $m$. In
this section, the exponents of the probability distribution as well as the
full distribution are calculated from a mean field analysis. The values of
$P(m)$ thus obtained are compared with the $P(m)$ for small values of
$\alpha$ obtained from Monte Carlo simulations in one dimension. From the
mean field analysis, we also calculate $P(m)$ for those values of $\alpha$
which are difficult to probe by Monte Carlo simulations due to the large
times required to reach the steady state. 

In the mean field approximation, all correlations are ignored by setting
all joint probability distributions to be the product of single point
distribution functions, $i.e.$, $P(m_i,m_j) = P(m_i) P(m_j)$. Under this
approximation, the $P(m)$'s evolve in time as
 \bea
 \frac{d P(m)}{dt}&=& - P(m)(m^{-\alpha}+ w +s'+ ws)  + w P(m+1) 
\nonumber \\*
 &+& s w P(m\!-\!1) + \sum_{a=1}^m \frac{P(a) P(m\!-\!a)}{a^{\alpha}},
m>0, \nonumber \\ \label{eq:35} \\
 \frac{d P_0}{dt} &=& -s' (1-s) -w s (1-s)+w P_1+s', \label{eq:36}
 \eea
 where $s'=\sum_{1}^{\infty} m^{-\alpha}$.  In the steady state, the time
derivatives vanish.  Multiplying by $e^{-p m}$ and summing $m$ from $1$ to
$\infty$, and eliminating $P_1$, we obtain
 \be
 Q =\frac{s Q' +w s (1-s) (1-e^{-p}) -s s'} {Q' -w s
 -s' -w +w e^p +w s e^{-p}} \label{eq:37}
 \ee
 where $Q=\sum_1^{\infty} P(m)e^{-p m}$, $Q'=\sum_1^{\infty} P(m) 
m^{-\alpha} e^{-p m}$ are generating functions.
 
The unknown quantities $s$ and $s'$ are determined by the two conditions
 \bea
 \left(Q'\right)_{p=0}&=&s' ,\label{eq:38}\\
 \left(\frac{d Q}{d p}\right)_{p=0} &=&-\rho . \label{eq:39}
 \eea
 Using the series expansion $Q = \sum_{n=0} \langle m^n \rangle (-p)^n/
n!$ and $Q' = \sum_{n=0} \langle m^{n -\alpha} \rangle (-p)^n/n!$, and
comparing terms order by order in $p$, we obtain relations between moments
of $P(m)$. From the term in $p^2$ we find
 \be
 \rho \langle m^{1-\alpha} \rangle = \rho w (1-s) - ws. 
 \label{eq:40}
 \ee
 Interestingly, Eq.~(\ref{eq:40}) is identical to the exact
Eq.~(\ref{eq:19})  in the $V\rightarrow \infty$ limit.  For $\alpha = 0$
and $\alpha = 1$ this yields the two results $s = {{\rho w - {\rho}^2}
\over {w(1+\rho)}}$ \cite{RM1} and $s = \frac{\rho w}{w+\rho w+\rho}$. 
 \begin{figure}
 \includegraphics[width=8.0cm]{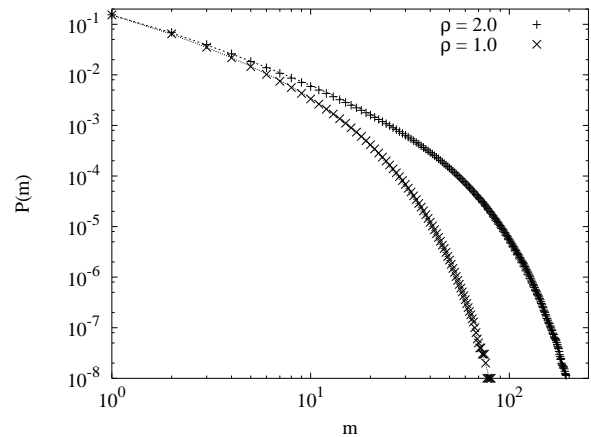}
 \caption{\label{fig4} $P(m)$ for $\alpha=1.0$ obtained from Monte Carlo
simulations (shown in symbols) are compared with the results from the mean
field analysis (shown as lines). The lattice size is $V=400$ and $w=1.0$
and we have used two densities $\rho = 1.0$ and $2.0$.}
 \end{figure}
 \begin{figure}
 \begin{minipage}{4.27cm}
 \includegraphics[width=4.27cm,height=4.27cm]{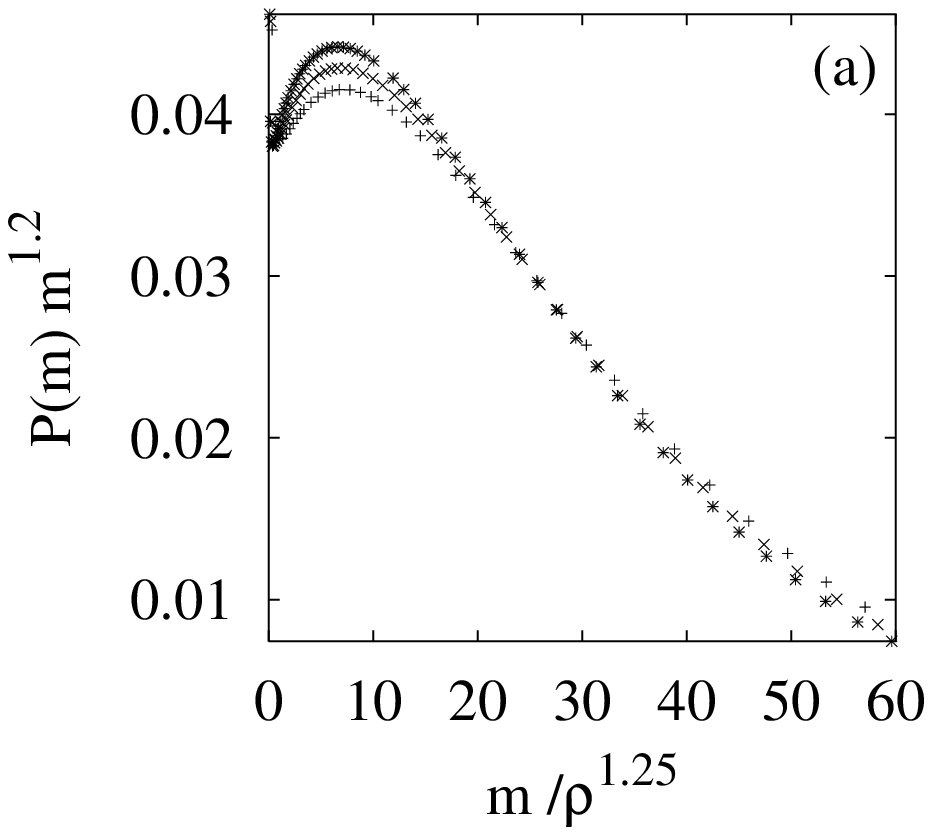}
 \end{minipage}
 \begin{minipage}{4.27cm}
 \includegraphics[width=4.27cm,height=4.27cm]{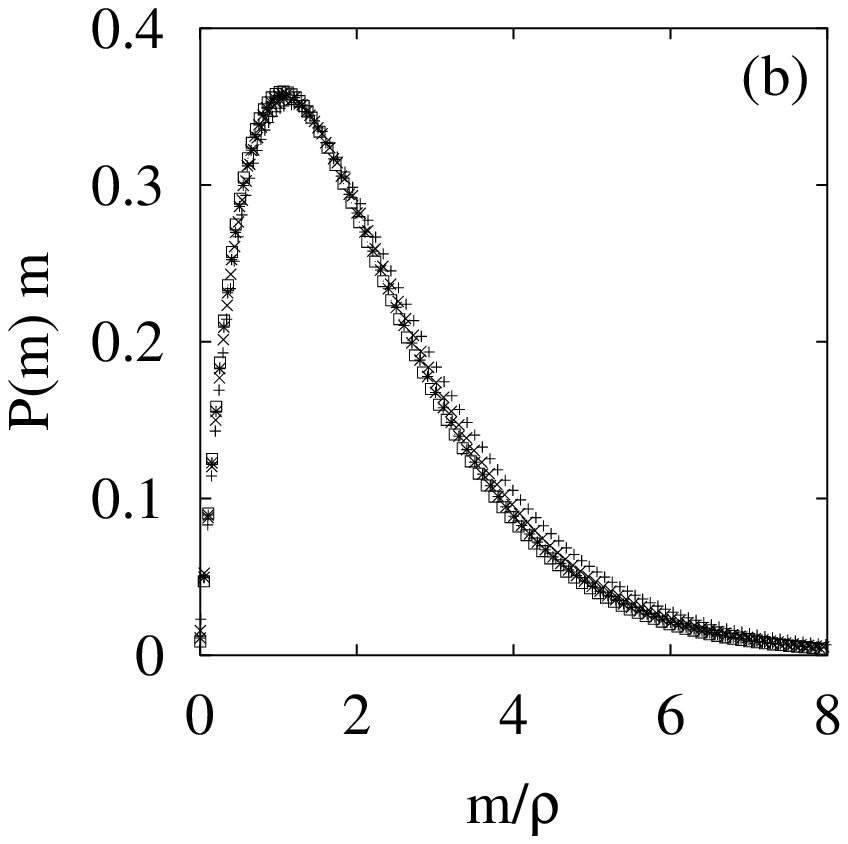}
 \end{minipage}
 \caption{\label{fig5} $P(m)$ obtained from the mean field analysis when
scaled as in Eq.~(\ref{eq:21}) with exponents as in Eqs.~(\ref{eq:31}) and
(\ref{eq:34}). (a) The curves are for $\alpha=1.6$, $w=0.1$ and for
densities $2.38$, $3.81$, and $6.69$.  (b) The curves are for
$\alpha=3.0$, $w=0.3$ and for densities $102.74$, $165.07$, $250.12$ and
$325.30$.  The scaling function $f_3(x)$ goes to zero for small $x$.}
 \end{figure}

Comparing the terms proportional to $p^3$, we obtain the relation
 \be
 \rho \langle m^{2-\alpha} \rangle = \langle m^2 \rangle {ws \over \rho} -
\rho w (1+s). 
 \label{eq:41}
 \ee
 Thus, for large $\rho$, $\langle m^2 \rangle \sim \rho^2 \langle
m^{2-\alpha} \rangle$. This provides us with a method for deriving the
exponents from the mean field equations. Assuming the scaling form
Eq.~(\ref{eq:21}) and using the exponent identity Eq.~(\ref{eq:24}), there
remain one independent exponent to calculate. Using the scaling form in
$\rho$, we obtain
 \be
 \phi (3-\tau')= 2+ \phi (3-\tau'-\alpha),
 \label{eq:42}
 \ee
 which immediately yields $\tau'=2-\alpha/2$ and $\phi = 2/\alpha$, the
same as in Eqs.~(\ref{eq:31}) and (\ref{eq:34}). 

We now calculate numerically the full $P(m)$ from the mean field
Eqs.~(\ref{eq:35}) and (\ref{eq:36}). If $s$ and $s'$ are known, the full
distribution $P(m)$ is known. We use this fact to determine the full
distribution numerically by the following procedure.  We fix $s$ and $s'$
at a certain initial value and calculate the resulting $P(m)$ and check
the consistency condition $s'=\sum P(m)/m^{\alpha}$. We tune $s'$ to
satisfy the above condition to an accuracy of $10^{-5}$, to determine
$P(m)$'s, and thus the density $\rho$. 

Using the above numerical method, $P(m)$ was calculated for various
$\alpha$'s. In Fig.~\ref{fig4}, we compare some of these mean field
results with $P(m)$ obtained using Monte-Carlo simulation, for large $V$.
The agreement is excellent, suggesting that the mean field results are a
very good approximation to the actual answer.

We now use the mean field results to probe $P(m)$ for values of $\alpha$
that cannot be studied easily by Monte Carlo simulations. In
Fig.~\ref{fig5}, we show the scaling plots for $\alpha=1.6$ and
$\alpha=3.0$. As mentioned in Sec.~\ref{sec4b}, the small $x$ behavior of
the scaling function has a different behavior for $\alpha<2$ and
$\alpha>2$. In the former case $f(x) \sim \mbox{const}$, while in the
latter case $f(x) \sim 0$ when $x\rightarrow 0$.

\section{Summary and conclusions}

In summary, we have studied the steady state of a system of aggregating
and fragmenting particles, with a mass dependent diffusion rate $D(m)\sim
m^{- \alpha}$ with $\alpha>0$.  We showed analytically that the
nonequilibrium phase transition which is known to exist for $\alpha = 0$,
vanishes when $\alpha > 0$. This is in agreement with the results of
extensive numerical simulations, through which we explored the dependence
both on system size and total mass.  Although no true infinite aggregate
forms in the thermodynamic limit, its imprint at high densities remains in
finite-sized systems in the form of an aggregate. Further, for the single
site mass distribution function, we obtained the exact scaling exponents
associated with its dependence on the mass, the density and the system
size. 

Our results give more credibility to the intuitive arguments presented in
\cite{RK} as to the circumstances in which one should expect to see a
nonequilibrium phase with an infinite aggregate, as occurs in the $\alpha
= 0$ case. We reproduce the argument here. In the model under
consideration, there are two competing processes: while the diffusion move
creates larger and larger masses by coagulation, the fragmentation move
tends to create smaller masses, as well as to inhibit the formation of
large masses. If the diffusion move was to be considered by itself, then a
cluster of size $l$ would be created in time of order $l^{2+\alpha}$. If
the fragmentation move was to be considered on its own, then a fluctuation
of order $l$ would be dissipated in time of the order $l^{2}$. This
exponent is known exactly because of the exact analogy \cite{MKB} in
one-dimension between an only-fragmentation model and the
Edwards-Wilkinson interface \cite{EW}. For $\alpha=0$, the two processes
are of similar strength and hence there is the possibility of a
transition. But for $\alpha>0$, the fragmentation process always dominates
and hence there is no aggregate phase. 
  
\section*{Acknowledgments}

We would like to thank S.~Coppersmith, S.~Redner, P.~L.~Krapivsky,
D.~Dhar, T.~Witten and S.~Krishnamurthy for helpful discussions, and
especially A.~Bray for his insightful suggestions on one of the limiting
cases. DD and BC were supported by NSF DMR-9815986. RR would like to thank
EPSRC, UK for financial support.

\appendix

\section{\label{appendix1}Exactly solvable limits}

In this appendix, we discuss the limiting cases of the model for which the
full probability distribution $P(m)$ can be calculated. 

\subsection{\label{appendix1a}$\alpha = \infty$}

In the limit $\alpha \rightarrow \infty$, the rate of diffusion becomes
equal to zero for all masses $m\ge 2$. The model then reduces to a zero
range process \cite{zrp} in which with rate $w$ unit mass can break off
from masses $m\geq 2$, while the unit mass can hop to a neighboring site
with rate $1+w$. It is then straightforward to verify that the steady
state probability distribution has a product form, \textit{i.e.},
 \be
 P(\ldots,m_1,m_2,\ldots) = \prod_i P(m_i),
 \label{eq:a1}
 \ee
 with
 \be
 P(m) = \cases{
 c \gamma^m & $m \geq 1$, \cr
 \frac{c (1+w)}{w} & $m=0$. \cr}
 \label{eq:a2}
 \ee
 The constants $c$ and $\gamma$ are fixed by the two constraints $\sum_m
P(m) =1$ and $\sum_m m P(m) = \rho$. Solving for $c$ and $\gamma$, we
obtain
 \bea
 c&=& \frac{w (1-s)}{1+w},\label{eq:a3}\\
 \gamma & =& \frac{s (w+1)}{w+s},\label{eq:a4}
 \eea
 with the site occupation probability $s$ being equal to
 \be
 s=\frac{\sqrt{w^2 (1+\rho)^2 + 4 \rho w} - w (1+\rho)}{2}. 
 \label{eq:a5}
 \ee
 In the limit $w\rightarrow \infty$, $s$ has the correct limit
$\rho/(1+\rho)$ (see Eq.~(\ref{eq:a9})). 

We would be interested in the form of $P(m,\rho)$ when $\rho \rightarrow
\infty$. Expanding $s$ in terms of $1/\rho$, we obtain,
 \be
 s=1- \frac{1+w}{w} \frac{1}{\rho}+ O(\frac{1}{\rho^2}). 
 \label{eq:a6}
 \ee
 In this limit
 \be
 P(m,\rho) \approx \frac{1}{\rho} e^{-m/\rho}, \quad \rho \rightarrow
\infty. 
 \label{eq:a7}
 \ee
 Thus
 \be
 P(m,\rho) =\frac{1}{m} f_\infty(\frac{m}{\rho}), \quad m,\rho \rightarrow
\infty,
 \label{eq:a8}
 \ee
 where the scaling function $f_\infty(x) \sim x $ when $x\rightarrow 0$. 
From Eq.~(\ref{eq:a8}), we see that $\tau=1$ for $\alpha=\infty$.

\subsection{\label{appendix1b}$w=\infty$}

In the limit $w=\infty$, the model reduces to a zero range process
\cite{zrp}. As in the $\alpha=\infty$ case, the steady state probability
distribution has a product form as in Eq.~(\ref{eq:a1}). $P(m)$ for this
limiting case was worked out in \cite{MKB}. For the sake of completeness,
we reproduce the final result,
 \be
 P(m) = \frac{1}{1+\rho} \left( \frac{\rho}{1+\rho}\right)^m, \quad
 m\geq 0. 
 \label{eq:a9}
 \ee

\subsection{\label{appendix1c}$w = 0$}

In this limit, masses diffuse and coagulate on contact.  Clearly, the
steady state is one in which the entire mass is clumped together into one
aggregate. For the $\alpha>0$ problem, this is the only limit in which an
aggregate forms which holds a finite fraction (here unity) of the total
mass at finite density.

\end{document}